\documentclass[12pt]{article}
\input{epsf.sty}
\usepackage{hyperref}
\usepackage[pdftex]{graphicx}
\usepackage{amsmath, amssymb}
\def\P{{\Bbb P}}

\def\phi{\varphi}

\def\s{\sigma}

\def\L{\Lambda}

\def\P{{\Phi}}

\def\T{\T}

\begin{document}
\title{ On the prevalence of non-Gibbsian states in mathematical physics}

\author{
Aernout C.D. van Enter
\footnote{ University of Groningen,Johann Bernoulli Institute of Mathematics and
Computing Science, Postbus 407, 9700 AK Groningen, The
Netherlands,
\newline
 \texttt{A.C.D.v.Enter@math.rug.nl}, 
\newline
\texttt{http://www.math.rug.nl/~aenter/ }}
\newline
}

\maketitle

\begin{abstract} 
Gibbs measures are the main object of study in equilibrium statistical 
mechanics, and are used in many other contexts, including dynamical systems 
and ergodic theory, and spatial statistics. However, 
in  a large  number of natural instances one encounters measures that are not 
of Gibbsian form. 
We present here a number of examples of such non-Gibbsian measures, and discuss 
some of the underlying mathematical and physical issues to which they 
gave rise. 

 \end{abstract}


 \smallskip

\vfill\eject

\section{Introduction}
\subsection{Gibbs  measures according to DLR}
Gibbs (or DLR) measures,  or Gibbs states, are the main objects in 
classical equilibrium statistical mechanics.
They were introduced in the sixties by Dobrushin, Lanford and Ruelle, as 
probability measures on systems of infinitely many particles (or spins) 
in infinite volume,
satisfying a set of consistent conditional probabilities for 
configurations in finite volumes, conditioned on external configurations.
This is expressed by the so-called DLR equations.  
These conditional probabilities   are of the Gibbsian form $Cst \  \exp-\beta H$, 
where the Hamiltonian $H$ describes the interactions between particles 
both inside the volume and between the volume and the outside.
In particular for classical lattice models, the theory of infinite-volume 
statistical mechanics  has developed in substantial detail, see e.g. \cite{Geo}.
Gibbs measures also play a role  in 
various other domains, such as Dynamical Systems, 
non-equilibrium theory,  Interacting Particle Systems, Euclidean Quantum 
Field Theory, ergodic theory, spatial statistics and pattern recognition.

For finite-range interactions, Gibbs measures satisfy a spatial Markov property,
for regular infinite-range potentials a weak form thereof, which goes by the 
names of the  `` almost Markov'' or ``quasilocality'' property. For discrete
bounded-spin models this property is equivalent with the conditional 
probabilities being continuous functions of the boundary conditions, 
in the product topology.

It is a nontrivial result that in fact equivalence holds: any measure whose 
conditional probabilities are quasilocal is  a Gibbs measure for a 
regular interaction, once it satisfies a natural nonnullness condition. 

\smallskip
\subsection{Effective descriptions and the Gibbs-non-Gibbs question}
In many parts of statistical physics use is made of effective interparticle 
Hamiltonians.
That is, one tries to describe a system in which one forgets about small-scale 
details, but that still can be described by a Hamiltonian, which contains only
properties of larger-scale entities. (E.g. an effective molecular Hamiltonian 
does not include properties of the constituent atoms, electrons or quarks,
but only -effective- forces between molecules).  
To make this notion mathematically precise,  that is, to decide if such an 
effective Hamiltonian exists,  the quasilocality property mentioned above 
needs to be checked for an appropriate measure (in the example above, 
that would be the measure restricted to  all the molecular degrees of 
freedom).  
However, it has turned out, initially rather surprisingly, 
that in many  natural examples this quasilocality 
property is violated, and no regular interaction can be found: 
the measure is {\it non-Gibbsian}. Often, the Gibbsian or non-Gibbsian character
of a measure depends on certain parameters of the problem under consideration, 
such as temperature, magnetic field, time, or  rescaling parameters,   
in an a priori non-obvious way. 

Examples of such contexts that occur in statistical mechanics are 
Renormalisation Group theory, the theory of disordered systems and the theory 
of stochastic dynamics (Interacting Particle Systems).

In Renormalisation Group  theory, a Renormalisation Group 
map is a kind of coarse-graining 
map.  One considers only a subset of coarse-grained or 
``renormalised'' objects (spins, fields), 
and then considers the restriction or projection of Gibbs states on those.   
In physical terms, one integrates out some short-range degrees of freedom. 
In probabilistic terms, one takes the  marginal of a probability 
measure on a  subset of random variables, often called the set of  
``block spins''.  
This renormalised measure then is 
supposed to be describable by a renormalised Hamiltonian. The philosophy of 
Renormalisation Group theory is based on studying
the properties of this 
map from  original  to  renormalised Hamiltonians 
in some appropriate space. 
The ultimate goal is to determine 
the fixed points of 
this map, together with their stability properties, and to relate 
them to the critical behaviour inside 
corresponding ``universality classes'', which are classes of physical 
systems with the same critical exponents.  
In this way, 
critical behaviour is expected to follow from the 
properties of certain Renormalisation Group maps. 
This paradigm is based on the assumption 
that such a map 
exists, in other words, that the renormalised measure {\em is} in fact
 a Gibbs measure.  

It is precisely this step which has turned out to be doubtful in a variety of 
circumstances. The  first clear indication that defining a well-behaved 
Renormalisation Group map might be problematical  was found by 
Griffiths and Pearce \cite{GP} and the nature of the problem was identified 
shortly after by Israel \cite{Isr2}. A first extensive analysis appeared in 
\cite{EFS}.  We will see the mechanism in the particularly simple 
example of the decimation transformation later on.
Although in a majority of cases the predictions of the 
Renormalisation Group approach about the 
nature of phase transitions and critical properties are not affected, in 
some cases, especially in the theory of first-order transitions, 
non-Gibbsianness results restricted and even excluded  
Renormalisation Group descriptions proposed in the physics literature.

Follow-up studies identified a variety of other occurrences of  
non-Gibbsian measures.  A direct generalisation of the above treatment of 
block-spin maps often works in considering 
single-site renormalisations, including dicretisations,  the so-called 
``fuzzy'' or ``amalgamation'' maps \cite{CU,EKO,Ver}.

Another example is provided by
low-temperature Gibbs measures, 
subjected to a high-temperature or infinite-temperature 
Glauber (stochastic spin-flip) 
dynamics. This  is an example from the area of interacting particle systems 
\cite{lig}, which models a fast heating 
procedure.  The initial Gibbs measure 
after some finite time can become non-Gibbsian. So instead of raising the 
temperature \cite{OP}, one may altogether 
lose the notion of effective  temperature 
\cite{EFHR}. 
The proofs of such non-Gibbsianness results are quite similar to the 
ones in a Renormalisation Group context, but with the distinction that one 
may consider now the 
marginal of a two-time (initial time and end time) system which is of a 
Gibbsian form. One can in fact go into  more detail, and perform a path-space 
analysis in which the whole dynamics is included \cite{EFHR2}. 

Yet another family of occurrences  of non-Gibbsian measures is  
in the theory  of disordered spin systems \cite{bov}. 
In such systems the Hamiltonians contains, next to the spin variables, 
disorder variables, e.g. occupation numbers or random fields. 
When a ``quenched''  disordered measure is non-Gibbsian, that  means that one 
cannot write it as an ``annealed'' measure, that is a Gibbs measure 
for an effective Hamiltonian. 
Physically,  in a quenched, fast cooled, system the disorder is frozen, while 
the spins equilibrate; in other words, the disorder variables are slow and the 
spins fast.  In  annealed, slowly cooled, systems the disorder variables 
equilibrate with the spins, and there is only one timescale, and there are no 
fast or slow variables.
Probabilistically, for  a quenched measure
the disorder variables are independent, identically distributed;  conditioned
on the disorder variables the spins are distributed according to
a Gibbs measure.   
Annealed measures are Gibbs measures on a product space of spin and 
disorder variables. The impossibility to write a quenched measure 
as an annealed measure is in contrast to what has been
proposed in the physics literature as the Morita approach \cite{EMSS, Kuh, Ku}
where one aims to compute a ``grand potential'', an effective interaction for 
a quenched measure, viewed as a Gibbs measure (an annealed one).

\section{Gibbs measures and non-Gibbsian measures}

\subsection{Notation and Definitions}
We will consider lattice spin systems with a single-spin space $\Omega_{0}$, on a lattice $Z^d$, and a configuration space $\Omega= {\Omega_{0}}^{Z^D}$.
We will, for simplicity, mainly consider  Ising models, 
for which $\Omega_0= \{-1,+1\}$. 
We will indicate the spin variables at site $i$ by 
$\sigma_i$, 
$\omega_i$, $\eta_i$, and 
similarly spin configurations in a volume $\Lambda$ 
by $\sigma_{\Lambda}$, 
$\omega_{\Lambda}$, $\eta_{\Lambda}$.
 
We will consider Gibbs measures, which are defined for absolutely summable 
interactions $\Phi$ via the DLR equations. An interaction $\Phi$ is a (translation-invariant) collection of functions $\Phi_X(\sigma_X)$. Each $\Phi_X$ 
describes an energy contribution in a finite subset $X$ of the lattice. 
 Absolute summability  means that 
$\sum_{0 \in X} ||\Phi_X|| < \infty$.  This implies that any finite change in an 
infinite-volume configuration only comes with a finite energy cost (or gain), 
uniformly in the external configuration. 
Such interactions form an interaction (Banach) space.
The DLR equations say that given an 
external configuration $\eta_{\Lambda^c}$, 
the probability (density) of configurations in 
a volume $\Lambda$ is given by the Gibbs expression 
\[
\mu_{\Lambda}^{\eta_{\Lambda^c}}(\sigma_{\Lambda}) = 
\frac{\exp - \beta H^{\Phi}_{\Lambda}(\sigma_{\Lambda}\eta_{\Lambda^c})}{Z_{\Lambda}^{\eta_{\Lambda^c}}}.
\]
where 
\[
H^{\Phi}_{\Lambda} =\sum_{A;\; A\cap \L\neq \emptyset}\P_A(\s_\L\eta_{\L^c})
\]
This should hold for all volumes $\Lambda$, internal configurations $\sigma_{\Lambda}$ and external configurations $\eta_{\Lambda^c}$.
The conditional probabilities given above have a continuous (in the product 
topology) version due to the summability of the interaction. This means that the
conditional expectation of any local observable cannot change much between 
two configurations which are identical in a sufficiently large environment, 
and are different only far away,
whatever the configuration in this finite environment is. Each such 
configuration is thus a point of continuity for each conditional expectation. 

It also turns out to be true that a measure having a continuous version of 
its conditional probabilities, and satisfying a nonnullness 
(or ``finite-energy'') condition, is a Gibbs measure for a reasonable 
interaction.  
The finite-energy condition for Gibbs measures follows immediately from 
the absolute summability. 
\noindent
See e.g. \cite{EFS,DEZ,Geo} for further background.
  
In the standard nearest-neighbour Ising model
we have 
\[
 -H_{\Lambda} = \sum_{< i,j > \in \Lambda}\sigma_i \sigma_j + \sum_{< i \in \Lambda, j \in \Lambda^c >} \sigma_i \eta_j .
\]

\subsection{Decimating the Ising model, a paradigmatic  example}

Decimation, in which one considers just a subset of the spins, is a 
conceptually easy example of a Renormalisation Group map.
Let us consider the even decimation of the two-dimensional Ising model, in which
$\sigma_{i,j}'= \sigma_{2i,2j}$. Thus we consider only a quarter of the spins, 
namely those on sites  with both coordinates
even. Those primed spins will be our renormalised or ``visible'' spins. 
If the original Gibbs measure is at low enough temperature, the primed 
measure defined by taking the marginal of this  measure on the primed spins 
is non-Gibbsian. 
Indeed, let us fix  all primed spins in a large box in
an alternating configuration. Then 
the other, ``invisible'' spins in the box 
don't feel any influence from them, due to 
cancellation effects. Thus the conditioned system of the invisible spins, 
forms a spin system on a lattice with periodic holes, a ``decorated'' lattice.  

Any configuration of the visible (renormalised) spins acts as a 
condition in a conditional probability of the invisible-spin system, 
conditioned on it. But it is the alternating configuration which will be the 
one that we will show to be responsible for  non-Gibbsian behaviour.

If in an annlulus outside the box all visible spins are plus (that is, they are 
pointing upwards), we have a 
plus-like  boundary condition, for {\em any}  condition of the invisible 
spins outside the annulus. Now let us unfix the visible spin at the origin. Then
this spin has a positive expectation, larger than some constant, 
uniformly in the size of the box. Making the visible spins minus outside the 
box produces a minus expectation. Thus the visible spin at the origin, 
conditioned on a large surrounding alternating configuration of visible spins 
has a large change in expectation, when one changes the configuration far away. 

Notice that the phase transition in the system of invisible 
spins gets translated in a nonlocal influence --action at a distance-- between 
the visible spins, violating the quasilocality condition for the measure on 
the visible spins. This renormalised measure thus is non-Gibbsian. 
The alternating configuration is a point of discontinuity of the spin at 
the origin, conditioned on (considered as a function of) the visible spins. 
This argument works if the temperature is low enough, as the decorated lattice 
has a strictly lower transition temperature than the original Ising model. 

Although there are other  choices possible than the alternating configuration, 
we expect that  in fact for most choices of the primed-spin configuration 
continuity holds.

It can be shown that renormalising different  Gibbs measures for the same 
interaction results  in the renormalised measures being all Gibbsian or all 
non-Gibbsian. 

By  similar arguments other decimated measures
become non-Gibbsian. This includes a finite number of decimations applied 
to  Ising models in dimension at least two
in a weak field at low temperatures, or (arbitrarily often)  
repeated decimations in zero field at low temperatures. 
In the zero-field case
the alternating configuration is neutral, in that it does not favour either the plus or the minus phase.  The Ising model with a small plus field,  
does not exhibit multiple phases, but
conditioning 
on a configuration which is predominantly minus can induce a phase transition. 
Thus the presence of a phase transition in the original system is neither 
necessary, nor sufficient, for the transformed measure to be non-Gibbsian.

\smallskip
On the other hand, at high temperatures, and also for 
decimation in strong fields, 
the transformed measures are Gibbsian. 
Thus, as Griffiths and Pearce \cite{GP} already noticed, one can define 
Renormalisation 
Group maps where one does not really need it, away from phase transitions (and 
even then not always). For mathematical details, see \cite{EFS}.  But even then, 
the Renormalisation Group map on the space of summable interactions has 
unexpected spectral properties, indicating that this space, although 
giving rise to proper Gibbs measures, is already too large 
to properly implement Renormalisation Group ideas in (see \cite{Yin}).
\bigskip


\subsection{Extensions I: Renormalisation, stochastic dynamics, 
discretisations} 

The occurrence of non-Gibbsian measures is actually quite widespread. 
Indeed, in a topological sense, they occur generically, for a residual set 
(that is,  a countable intersection of  dense open sets) in the set of 
probability measures 
\cite{Isr3}.

Similar results as proven above for decimation
can be proven for a variety of Renormalisation Group 
transformations. 
For example, one can prove non-Gibbsianness for Ising models subjected to 
majority-rule transformations (in which a renormalised spin equals the sign of 
the majority of the spins in a block) at 
low temperatures in any external field, various random versions thereof (the 
Kadanoff transformations), etc.

Beyond Renormalisation Group transformations, similar results hold  also for 
evolved Ising systems, under a high-temperature Glauber (stochastic spin-flip)  
dynamics. 
Starting from a low-temperature Gibbs measure in the phase-transition regime, 
for a short time the evolved measure is Gibbsian, but at larger 
times it becomes non-Gibbsian, and then it stays so for any finite time in this 
transient, nonstationary,  regime. 
This is true although the measure converges exponentially fast to a 
very well-behaved high-temperature Gibbs measure. 
Other sources of non-Gibbsian measures are  
single-site coarse-grainings. 
In the dynamical case  the visible spins are evolved spins, and the 
invisible ones the initial spins.  
For single-site coarse-graining (fuzzy \cite{Ver} or amalgamation \cite{CU}) 
maps, fine details become invisible, 
and one can only observe coarser details, the fuzzy, or amalgamated, spins.

In all these examples, the presence of a transition in the invisible spins, 
conditioned on some special configuration of the visible spins,  gets 
translated into the fact that  this special configuration is 
a point of discontinuity ( a ``bad point''). 
If for no possible conditioning a phase transition occurs,  the 
transformed or evolved measure is a Gibbs measure. This typically happens if the
transformation is close to unity. Examples of Gibbsian regimes 
are  very-short-time evolutions, or 
very fine discretisations for initial systems that are at not too low 
temperatures.   

Let me emphasize that the absence or presence of these transitions is for  
conditioned systems, and {\em not} for the original, untransformed system, which
may or may not be in a phase transition regime.  

\subsection{ Extensions II: Trees and Mean-Field theory. Path approach }
Related results can be proven in a mean-field setting, in this case, the 
(dis)continuity to be investigated of, for example, a spin expectation,
is not any more of that of a function of the external configurations 
in the product topology. Rather, the conditional 
expectation value of a spin is seen as   
a function of some order parameter, such as a magnetisation.
This approach has especially been pioneered by C. K\"ulske, see e.g. 
\cite{KuLe, Len1}. 

In the above context, the ``bad points'' are exceptional, that 
is they have measure zero. In other situations, in particular in the Random 
Field Ising model, and also for evolved unstable 
Gibbs measures on trees, it can even 
happen that {\em almost all} or {\em all} configurations 
become bad \cite{EEIK,KLR}. 
Gibbs measures on trees differ from those on lattices in that, due to the 
large boundary terms, at low temperatures one can have metastable and even 
unstable homogeneous Gibbs measures, corresponding to different types of 
solutions of a self-consistency equation. In this sense they violate the 
variational principle that says that all Gibbs measures minimise a free energy 
density.  Due to this, the Gibbsian and non-Gibbsian properties of the evolved 
measures can be very different for different initial Gibbs measures for 
the same initial interaction.

Recently, in the dynamical Gibbs-non-Gibbs transitions a more refined analysis 
has led to the identification of bad objects (bad points or bad measures)  
as points or measures which can have different, competing, histories. A 
large-deviation analysis on the level of 
trajectories in a space of paths 
then becomes required. The corresponding rate functions are sums of
an initial rate function, and a dynamical rate function, which can be computed 
as a particular Lagrangian by the methods developed in \cite{FK}. For these 
developments we refer to \cite{EFHR2,EK,RW,FHM}. 
A bad value of the magnetisation 
then would be one which can have two quite 
different origins, starting from either a positive or a negative value, for 
example.

\subsection{Further generalizations. Other sources of non-Gibbsianness.}

The above description has been mostly about discrete-spin models, but extensions
to continuous, possibly   unbounded, spin systems also exist. 
In the bounded-spin case of vector models, the case which  has in particular 
been studied is that of stochastic time evolutions, see e.g.  \cite{EKOR}.
Continuous-spin systems have been studied  either be subjected to single-site
or weakly interacting  diffusions, or  (as mentioned before) to discretisation. 
In the unbounded-spin case, the notion of what is a Gibbs measure for a ``decent'' interaction becomes a bit more arbitrary. For some of the literature  on 
this issue, see \cite{DR, KuRe, RRR}.
In the case of discretisation of vector spins one determines the angle of an 
XY (vector) spin up to finite precision, obtaining a ``visible'' 
clock-spin measure, 
which in the Gibbsian situation has a summable clock-spin interaction, but at 
very low temperatures becomes non-Gibbsian \cite{EKO}.

Other examples of non-Gibbsian measures abound, including   
random-cluster (Fortuin-Kasteleyn)  
measures, invariant measures for stochastic evolutions,  g-measures, which 
satisfy a one-sided version of the continuity (Gibbs) property of  their 
conditional probabilities, lower-dimensional projections of Gibbs measures, 
sign-fields of massless Gaussians.... 
See e.g. \cite{FGM} and for earlier results  \cite{DEZ, EFS} or the special 
Vol 10(3) of the journal ``Markov Processes and Related Fields''. 
Next to a violation of  the quasilocality  property, another way of proving 
non-Gibbsianness which  works in some of the above cases,
is showing either anomalous 
large-deviation properties, or a violation of the non-nullness (or 
finite-energy) condition. 

Another, as yet unexplored, direction is about 
quantum statistical mechanical systems. In this 
case one is still looking for a characterisation of Gibbs or KMS states which 
can actually be checked in examples. Conditional probabilities have no analogue
in a quantum context, which makes the above classical analysis not applicable.

\section{Conclusions}
Although  a variety of examples of non-Gibbsian measures have by now been 
discovered,  the significance of this fact still appears somewhat controversial.

Mathematically, the phenomenon seems quite widespread, and we have developed 
a fairly systematic approach to handle a lot of examples, many of which are 
measures showing up in natural circumstances.

One response has been to try to make non-Gibbsian measures `` as Gibbsian as 
possible'', by weakening the definition of what a Gibbs measure is. This 
approach, which was suggested by R.L. Dobrushin,  has led to the notions of 
almost, weak and intuitively weak Gibbs measures \cite{MMR,DS,EV}. As a 
warning, 
it should be noted that the quenched Random Field Ising measure, which can be 
shown to be weakly Gibbsian, (that is, one can define a Hamiltonian which is 
defined almost everywhere with respect to
 this measure), violates the variational principle 
\cite{KLR}.  This implies that
 measures from  these classes can be substantially less 
well-behaved than regular  Gibbs measures.  

Physically, non-Gibbsianness reflects the presence of 
some nonlocal correlations, describable by
interactions that have an  extremely  long range. They are not even summable 
and  represent some ``actions at a distance''.  
In the theory of the Renormalisation Group, as applied to critical phenomena,  
the appearance of 
long-range interactions often leads to these interactions belonging to a 
different universality class, even if they are summable. Hence  there is 
serious cause for concern if terms appear which are even worse.  Ideally, a 
Renormalisation Group map would act on a subspace of interactions within the 
same universality class. 

The fact that a proposed algorithm is mathematically ill-defined may or may not 
invalidate results which are obtained by approximate methods. However, for a 
mathematical physicist to develop a systematic understanding when and when not 
to trust renormalisation-group folklore remains a big challenge. 
Similar questions arise if one tries to introduce effective temperatures, 
or effective potentials of the Morita (quenched-as-annealed) type. 

 If one can prove the Gibbsianness of a measure, one can in principle  trust 
numerical approximations, and hopefully 
obtain  some proper error bounds. However, even that appears to be a much 
harder problem than one would a priori expect \cite{Ken}.



\section{Acknowledgements} I have discussed about and 
worked on   non-Gibbsian measures with many colleagues, whom I 
sincerely thank for all they have taught me. In particular I wish to 
thank Roberto Fern\'andez for our longstanding and always pleasant 
collaboration. I thank Valentin Zagrebnov for inviting me to write this 
review and Henk Broer, Roberto Fern\'andez and Siamak Taati for 
helpful suggestions on the manuscript.


\begin{thebibliography}{99}


\bibitem{bov} A.~Bovier: Statistical mechanics of disordered systems.
Cambridge University Press, Cambridge (2006).

\bibitem{CU} J.-R.~Chazottes and E.~Ugalde: On the preservation of 
Gibbsianness under symbol amalgamation. 
In: Entropy of hidden Markov processes 
and connections to Dynamical Systems. Eds. B.~Marcus, K.~Petersen and 
T.~Weissman. LMS Lecture Notes 385, 72-97 (2011).


\bibitem{DR}D.~Dereudre, S.~Roelly: Propagation of Gibbsianness for 
infinite-dimensional gradient Brownian diffusions. J. Stat. Phys. 121, 511-551 (2005).

\bibitem{DS}  R.L. Dobrushin and S.B. Shlosman: Non-Gibbsian states and their Gibbs description. Comm. Math.Phys. 200, 125--179 (1999).





\bibitem{EEIK} A.C.D.~van Enter, V. Ermolaev, G. Iacobelli and C. K\"ulske:
Gibbs-non-Gibbs properties for evolving Ising models on trees. Ann. 
Inst. H. Poincar\'e, to appear. arXiv 1009.2952 (2010).

\bibitem{EFS} A.C.D.~van Enter, R.~Fern\'andez, A.D.~Sokal: Regularity properties 
and pathologies of position-space renormalization-group transformations: Scope and 
limitations of Gibbsian theory. J. Stat. Phys. 72, 879-1167 (1993).

\bibitem{EFHR} A.C.D.~van Enter, R.~Fern\'andez, F.~den Hollander, F.~Redig: 
Possible Loss and recovery of Gibbsianness during the stochastic evolution of 
Gibbs Measures. Comm. Math. Phys. 226, 101-130 (2002).

\bibitem{EFHR2} A.C.D.~van Enter, R.~Fern\'andez, F.~den Hollander, F.~Redig:
A large-deviation view on Dynamical Gibbs-non-Gibbs transitions. 
Moscow Math. J. 10, 687-711 (2010).

\bibitem{EKO} A.C.D. van Enter, C. K\"ulske and A.A. Opoku: Discrete 
approximations for vector spin models. J. Phys. A, Math. Theor. 44, 475002 (2011). 

\bibitem{EKOR} A.C.D. van Enter, C. K\"ulske,  A.A. Opoku and W.M. Ruszel: 
Gibbs-non-Gibbs properties for n-vector lattice and mean-field models. Braz. J. Prob. Stat. 24, 226-255 (2010).

\bibitem{EMSS} A.C.D. van Enter, C. Maes, R.H. Schonmann and S.B. Shlosman:
The Griffiths singularity random field. In {\em On Dobrushin's way. From probability to statistical mechanics.} 51-58, Am. Math. Soc.(2000).
 
\bibitem{EV} A.C.D.~van Enter and E. A. Verbitskiy: On the variational principle for generalized Gibbs measures.
Markov Proc. Rel. Fields 10, 411-434 (2004).
 
\bibitem{EK} V. Ermolaev and C. K\"ulske: Low-temperature dynamics 
of the Curie-Weiss model: Periodic orbits, multiple histories and loss of 
Gibbsianness. J. Stat. Phys. 141, 727--756 (2010).
 





 
 



\bibitem{FK} J. Feng and T.G. Kurtz: Large deviations for stochastic processes. 
AMS, Math Surveys and Monographs 131, Providence, Rhode Island (2006).



\bibitem{DEZ} R.~Fern\'andez: Gibbsianness and non-Gibbsianness in lattice 
random fields. Mathematical Statistical Physics, Les Houches, 
LXXXIII, Elsevier, Eds. A. Bovier, J. Dalibard, 
F. Dunlop, A.C.D. van Enter and F. den Hollander (2005).

\bibitem{FGM}
R. Fern\'andez, S. Gallo and G. Maillard: 
Regular g-measures are not always Gibbsian. El.Comm. Prob. 16, 732--740 (2011). 

\bibitem{FHM}
R. Fern\'andez, F. den Hollander and J. Mart\'inez: 
Variational description of dynamical Gibbs-non-Gibbs transitions for the Curie-Weiss model. arXiv 1202.4205 (2012). 





\bibitem{Geo} H.-O.~Georgii: Gibbs measures and phase transitions. Volume 9 of de Gruyter Studies
in Mathematics. Walter de Gruyter Co., Berlin, 2nd edition 
(2011).





\bibitem{GP} R. B.~Griffiths, P. A.~Pearce: Mathematical properties of position-
space renormalization-group transformations. J. Stat. Phys. 20, 499-545 (1979).








\bibitem{Isr2} R.B.~ Israel: 
Banach algebras and Kadanoff transformations.   
In: {\em Random Fields, Esztergom Proceedings},  North Holland, 593--606 (1981). 

\bibitem{Isr3} R.B.~ Israel: 
Some generic results  in  Mathematical Physics. 
Markov Proc. Rel. Fields 10 517--521 (2004).

\bibitem{Ken} T.G. Kennedy: 
Renormalization group maps for Ising models in lattice gas variables.
J. Stat. Phys. 140, 409--426 (2010).

\bibitem{Kuh} R. K\"uhn: 
Gibbs versus non-Gibbs in the equilibrium ensemble approach to 
disordered systems.
Markov Proc. Rel. Fields 10 523--545 (2004).

\bibitem{KLR}
C. K\"ulske, A. LeNy and F. Redig: Relative entropy and variational properties 
of generalized Gibbs measures. Ann. Prob. 32, 1691--1726 (2004).

\bibitem{Ku}C.~K\"ulske: 
How non-Gibbsianness helps a metastable Morita minimizer to provide a 
stable free energy. Mark. Proc. Rel. Fields 10, 547--564 (2004). 










\bibitem{KuLe} C.~K\"ulske, A.~Le Ny: Spin-flip dynamics of the Curie-Weiss 
model: Loss of Gibbsianness with possibly broken symmetry. Comm. Math. Phys. 
271, 431-454 (2007). 

\bibitem{KuRe} C.~K\"ulske, F.~Redig: Loss without recovery of Gibbsianness 
during diffusion of continuous spins. Prob. Theory Relat. Fields 135, 
428-456 (2006).

\bibitem{lig} T.~M. Liggett: Interacting Particle Systems. 
Springer Classics in Mathematics, reprint from the 1985 Edition (2005).

\bibitem{MMR} C.~Maes, A. van Moffaert and F.~Redig
Almost Gibbsian versus weakly Gibbsian measures. Stoch. Proc. Appl. 
79, 1-15 (1999).




\bibitem{Len1} A.~Le Ny:
Gibbsian Description of Mean-Field Models.
In: In and Out of Equilibrium, Eds. V.Sidoravicius, M.E. Vares, Birkh\"auser, 
Progress in Probability, vol 60, 463-480 (2008). 



%





 
\bibitem{OP} A. Petri, M.J. de Oliveira:
Temperature in out-of-equilibrium lattice systems.
Int. J. Mod. Phys.  C 17, 1703-1715 (2006).


%
%





\bibitem{RRR} F. Redig, S. Roelly and W.M. Ruszel: Short-time Gibbsianness for infinite-dimensional diffusions 
with space-time interaction. J. Stat. Phys. 138, 1124--1144 (2010). 



 
 
 
  


\bibitem{RW}
F. Redig and F. Wang:
Gibbs-non-Gibbs transitions via large deviations: computable examples.
arXiv 1202.4343 (2012).

 


%



\bibitem{Ver} E.A. Verbitskiy: Variational principle for fuzzy Gibbs measures, Moscow Math. J. 10, 811-829 (2010). 

\bibitem{Yin} M.Yin: Spectral properties of the renormalization group at 
infinite temperature. Comm . Math. Phys. 304, 175--186 (2011). 






\end{thebibliography}
\end{document}